# Observations of the Abundances of Secondary Galactic Cosmic Rays from Z=5-28 Between ~10-200 MeV/nuc Beyond the Heliopause by Voyager – Some Unexpected Anomalies and Their Interpretation Using a LBM for Galactic Propagation

W.R. Webber

New Mexico State University, Astronomy Department, Las Cruces, NM  88003, USA

**ABSTRACT**

Voyager observations for over 3 years beyond the heliopause have started to define features of the low energy ($\leq$ 100 MeV/nuc) cosmic ray secondary nuclei that have a zero or negligibly small source component. As an example, the abundance of B between about 7-15 MeV/nuc is unexpectedly large, greater than the prediction of a LBM by 2-3 $\sigma$ in the measurement and cross section error. On the other hand, for several other heavier secondary nuclei with a low source abundance such as F and Z=17-19 and 21-23 nuclei, in the corresponding energy channels between about 10-20 MeV/nuc, zero nuclei have been observed. The same LBM calculations would predict about 6-7 events for the sum of these three groups of nuclei. The B observed intensities could be more closely matched by considering a nested LBM with ~0.3-1.0 g/cm$^2$ of matter near the cosmic ray sources, essentially a "source" component of B. This nested LBM calculation, if extended to the production of the above groups of secondaries, would however predict a total 8-12 events thus leading to an even larger discrepancy with the zero events that are observed. The measurements of heavier secondaries at low energies therefore make it very problematical that a nested LBM with more than a few 0.1 g/cm$^2$ or more of matter near the sources could be the source of the large B abundance seen by Voyager. The large abundance of B at the lowest energies is best understood, so far, as the result of production mostly from a matter traversal of ~10 g/cm$^2$ in the galaxy, possibly in combination with uncertainties in the data and cross sections at these low energies.

**Introduction**

After ~3 years of observations beyond the heliopause the Voyager CRS instrument has now obtained reasonable statistics to study the abundance of secondary cosmic ray nuclei, those produced during interstellar propagation, at energies below ~100 MeV/nuc. For the most abundant of these nuclei, Boron and Nitrogen, the Voyager data abundances have been presented by Cummings, et al., 2015, 2016. Webber, 2015, has summarized and interpreted the measurements of these two nuclei in terms of a LBM for galactic propagation in some detail. For N nuclei the observed intensities as a function of energy are consistent with a simple LBM propagation where the path length below 1.0 GV decreases from ~10 g/cm$^2$ at ~100 MeV/nuc and above, to ~7 g/cm$^2$ at 12 MeV/nuc. The source abundance of N is determined from these calculations to be from 4-6% of $^{16}$O, similar to earlier Voyager observations within the heliosphere (Webber, et al., 1996), and observations near the Earth (Krombel and Wiedenbeck, 1988) and equal to 0.33-0.5 of the measured solar abundance ratio of these two nuclei. At the lowest energies the LBM calculations show, in fact, that most of the N is from the source component, so that future observations with improved statistics should improve the accuracy of this abundance ratio.

For B the Voyager observations are also consistent with the above model and the same path length used for N at energies greater than ~30 MeV/nuc. However, at lower energies and particularly between ~7-15 MeV/nuc there a is 2-3$\sigma$ excess of Boron in the Voyager observations above the predictions from a LBM according to Webber, 2015.

In this paper we will discuss the low source abundance nuclei with Z=9-28 that are also measured at Voyager and compare these measurements with those for B and N nuclei, and discuss the significance of the measurements of these higher Z secondary nuclei with regard to the galactic propagation at low energies.

**The Data**

In this paper we examine the abundances and intensities of several groups of secondary nuclei where, in some cases, interstellar production is important but not dominant and in some other cases where interstellar production is the dominant source of these nuclei, similar to the situations for N and B nuclei, respectively. The original intensities measured for each nucleus at Voyager have been presented in Cummings, et al., 2015, 2016. Here we have grouped these nuclei to include N, Na+Al and Ca+Cr+Ni nuclei as shown in Figures 1, 2 and 3. These nuclei are all expected to have a significant source component along with a significant interstellar component, similar to the situation for N, and are summed together because of statistical considerations. The spectra of B, Fl and Z = 17-19 and 21-23 nuclei, where the source component itself is now small and the interstellar production is dominant, similar to B, are shown in Figures 4, 5, 6 and 7. We will compare the measured spectra of these components with the propagation calculations for the latest parameters used in the LBM.

**A Discussion of the Spectra**

N, shown in Figure 1, has a source component 6.3% of O, which is significant mainly at energies $\leq$ 100 MeV/nuc. The Z=11+13 and 20-24-28 nuclei spectra shown in Figures 2 and 3 have estimated source components that are similar to N in terms of the source fraction of their principal progenitor nuclei, e.g., N=6.3% of O, Na=4.4% of Mg, Al=9.0% of Si, Ca=7.2% of Fe, Cr=3.2% of Fe and Ni=5.8% of Fe (these are the source abundances used in the propagation calculations for these nuclei shown in Figures 1 through 7).

The spectra for B, Fl, Z ~17-19 and 21-23 are shown in Figures 4, 5, 6 and 7. The spectra of these nuclei are plotted as though they have one count in the lowest energy interval even though they all have, in fact, zero counts through 06/30/15, compared to 34 counts for Boron in Figure 4 for the same time period. For Z=17-19 and 21-23 nuclei the source abundances used are 1.7% (for Ar) and 0.3 % (for Z=21-23) of Fe respectively.

The choice of the parameters for the LBM calculations is discussed in earlier companion papers by Webber, 2015 and Webber and Villa, 2016. Basically the path length, $\lambda$, is taken to be $\sim\beta\ P^{-0.45}$ above some rigidity, $P_0$, and $\sim\beta\ P^{1.0}$ below $P_0$.

Turning first to the intensity ratios for the somewhat more abundant secondary source nuclei such as N we find that, for the Z=11+13 nuclei spectra shown in Figure 2 and the Z=20+24+28 nuclei spectra which are shown in Figure 3, the difference between the LBM predictions using a path length $\lambda = 22.3\ \beta\ P^{-0.45}$ above 0.316 GV and $\lambda = 39.7\ \beta^{1.0}$ below 0.316 GV and the V1 measurements at all energies are not significant. At ~100 MeV/nuc the accuracy of the comparison is $\pm$ 10%. So they fit a consistent pattern. That is, all three of these spectra, which contain both a significant secondary and source nuclei component, are consistent with each other and are all consistent with a LBM calculation when $P_0$=0.316.

In the original calculations by Webber and Higbie in 2009, made before V1 crossed into interstellar space, $P_0$ was taken to be ~1.0 GV, and the fit to the LIS H and He spectrum later observed at V1 after 2012 was excellent. Since then it has been found that the LIS electron spectrum between about 3 and 70 MeV actually measured at V1 has a slope ~1.30 (Cummings, et al., 2016). This is greatly different than the slope of ~3.1 for the electron spectrum observed above ~10 GeV (Adriani, et al., 2013). This difference can be explained in part by a change in the rigidity dependence of the diffusion coefficient. But to match the high energy electron spectrum intensities measured by PAMELA (Adriani, et al., 2013) and simultaneously the V1 electron intensities using a constant spectral slope for the source spectrum of electrons acquires a break in the rigidity dependence below 1 GV. In fact, this new V1 electron data specifies a unique break at about $P_0$=0.316 GV. This implies that the high rigidity path length $\lambda=22.7\ \beta\ P^{-0.45}$ extends to lower rigidities than originally assumed, producing a much larger path length at low rigidities resulting in a larger production of secondary nuclei which is evident in all of the calculations shown in Figures 1-7.

Turning now to the zero or low source abundance nuclei in Figures 4-7 we find that to make the Voyager F nuclei measurements in Figure 5 consistent with the updated LBM propagation calculations would require ~1 F count and would also require about 2.5-3.5 counts each for Z=17-19 and 21-23 nuclei in Figures 6 and 7 in the lowest energy channel. So in this case the observed zero counts for each of the above nuclei in the lowest energy channels are not consistent with LBM calculations and are not consistent with the actual B measurements either.

**The “High” B Abundance and a Nested LBM**

The observed high B abundance thus appears to be an anomaly in the abundances of the low source abundance cosmic ray nuclei with Z<23 which are under abundant relative to both the B observations and the LBM calculations. If this high B abundance is due to excess B production near the source in as much as ~1 g/cm$^2$ of additional matter as in a nested LBM (e.g., Cowsik, Burch and Madziwa-Nussinov, 2014), then this “source” cannot be contributing to the production of F and Z=17-19 and 21-23 nuclei otherwise these nuclei would be even more under abundant at the lower energies as observed at Voyager.

Also the intensities expected for the Na+Al and Ca+Cr+Ni nuclei spectra shown by the blue curves for the NLBM in Figures 2 and 3, which would be, as a result, much higher than those actually measured for these nuclei, and this would also seem to rule out a simple nested LBM with as much as 1 g/cm$^2$, as the source for the excess B production.

**Summary and Conclusions**

This study of the intensities of low source abundance secondary nuclei below 100 MeV/nuc as measured by Voyager for cosmic ray nuclei with Z<26 appears to have reached two significant conclusions after nearly 3 years of data gathering.

First, at energies ~100 MeV/nuc and below, the abundances of the secondary nuclei with a significant source abundance which is in the range ~2-8% of the source abundance of Si, e.g., N, Na, Al, Ca, Cr and Ni, have spectra and intensities that are well predicted by a simple LBM where $\lambda$=22.3 x $\beta$ P$^{-0.45}$ above 0.316 GV and = 39.7 x $\beta^{1.0}$ for P < 0.316 GV.

Second, for nuclei with a low or zero source abundance such as B, F and the charge groups Z=17-19 and 21-23 nuclei, we have the situation where B is considerably over abundant, and all of the other nuclei have zero events, which is below the predictions of the LBM and the expectations from the measured B abundance. If one tries then to explain this high B abundance in terms of a nested LBM in which ~1 g/cm$^2$ of matter near the source produces the excess B as a "source" component, the predicted intensities of all the other secondary components are also increased by this excess production as well and do not fit the V1 data as seen in Figures 5-7.

So, to come the closest to explaining the Voyager B data in a simple LBM scenario, requires a large matter path length of at least 7.5 g/cm$^2$ at energies $\leq$30 MeV/nuc or possibly other modifications of the LBM including still a Nested LBM with a "source" corresponding to much less than 1 g/cm$^3$, perhaps as small as 0.1 g/cm$^2$ of local matter.

Recent calculations (Alosio, Blasi and Serpico, 2015) have shown that 0.1-0.2 g/cm$^2$ of matter is a possible expectation for cosmic rays escaping from the environment of supernova. In this case a "separate" source for the lowest energy B nuclei is still an alternative. An indication of the problems trying to explain the B abundance has occurred earlier in conjunction with ACE data (Davis, et al., 2000). This work considered the possibility of a local Carbon source for the production of the excess B that was observed by the ACE spacecraft.

As for the nuclei, F, Z=17-19 and 21-23, the lack of counts in the lowest energy channels is disturbing and outside of normal statistical uncertainties. At ~100 MeV/nuc the measured intensities of these nuclei are consistent with the LBM calculations to within $\pm$ 10%.

**Acknowledgements:** The author appreciates the efforts of all of the Voyager CRS team, Ed Stone, Alan Cummings, Nand Lal and Bryant Heikkila. It is a pleasure to work with them as we try to understand and interpret the galactic cosmic ray spectra that had previously been hidden from us by the solar modulation. We also thank JPL for their continuing support of this program for over 40 years. This work would not have been possible without the assistance of Tina Villa.

## FIGURE CAPTIONS

**Figure 1:** Spectrum of N measured by Voyager 1 (red points). Calculations are for a LBM with $\lambda$ = 22.3 $\beta$ $P^{-0-45}$ above 0.316 GV and $\lambda$ = 39.7 $\beta^{1.0}$ at lower rigidities. Total source abundance of $^{14}N$ and $^{15}N$ is 6.3% of $^{16}O$.

**Figure 2:** Combined spectra of Na and Al nuclei in cosmic rays measured by Voyager 1 (red points). The source abundances used in the LBM calculations are Na = 40.5 and Al 85 x Si, where Si = 1000. Also shown in blue is the Na+Al spectrum calculated using the same program but adding the production of "source" secondaries produced by 1 g/cm$^2$ of matter traversed near the source (NLBM) (blue line).

**Figure 3:** Combined spectra of Ca, Cr and Ni nuclei in cosmic rays measured by V1. Calculations are for the LBM noted above. The source abundances are Ca = 67.5, Cr = 28 and Ni = 55.7 x Si, where Si = 1000. Also shown is the calculation of extra secondaries from a nested LBM as noted in the Figure 2 caption (blue line).

**Figure 4:** Same as Figure 1 but for B.

**Figure 5:** Same as Figure 4. The spectra are for the F nuclei. The observation of zero counts in the lowest energy channel is placed at an intensity equivalent to 1 count. The green line, B, is the spectrum observed for B nuclei measured at Voyager normalized to the F intensity measurements at Voyager at 115 MeV/nuc.

**Figure 6:** Same as Figure 5, but the data is for Z = 17-19 nuclei. For $Ar_s$ (17 x Si=1000). The nested LBM calculation starts with source abundances derived from the passage of cosmic rays through ~1 g/cm$^2$ of matter near the source (blue line).

**Figure 7:** Same as Figure 6, but the data is for Z = 21-23 nuclei.

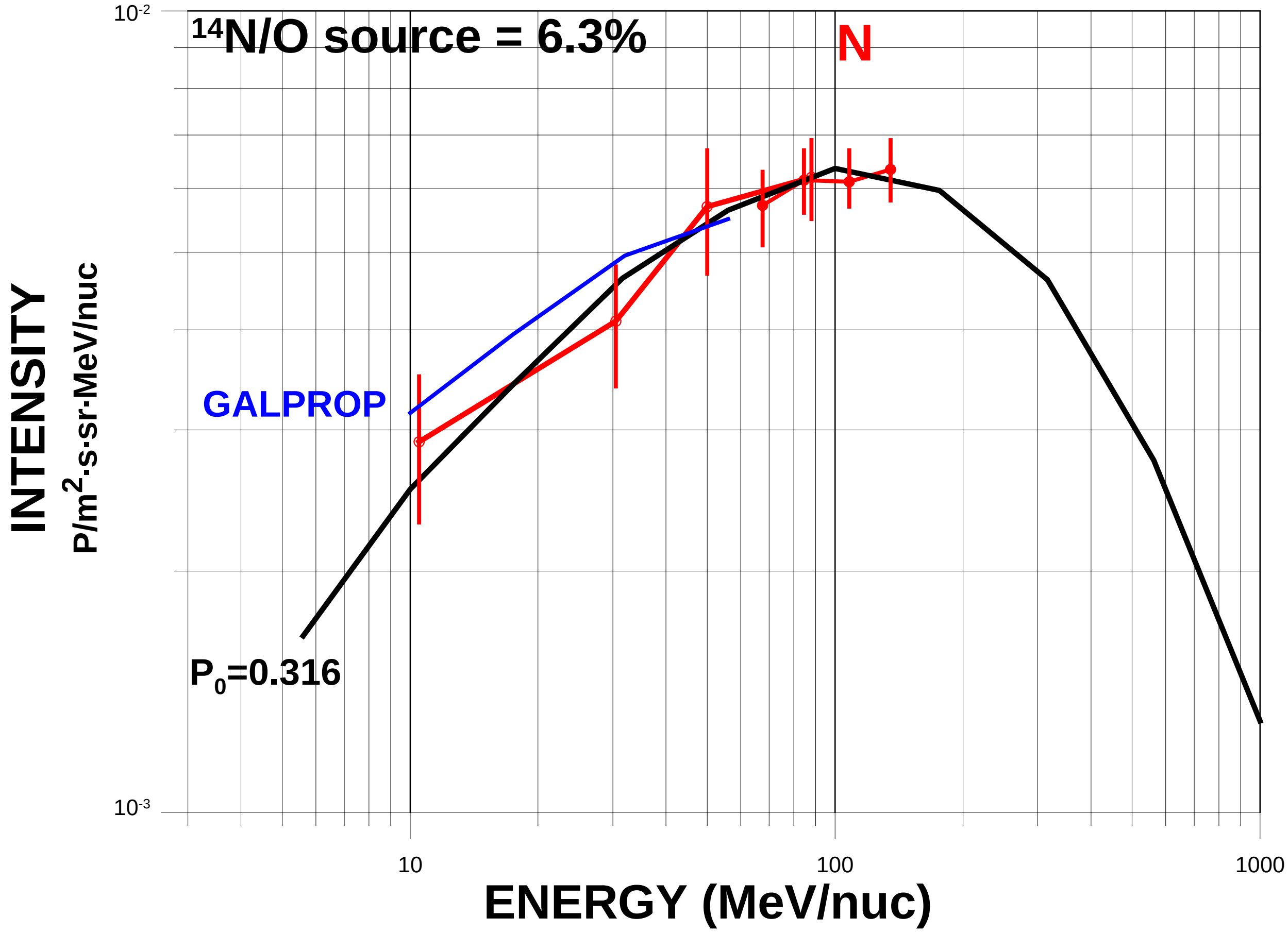

14N/O source = 6.3%
N
GALPROP
P0=0.316
INTENSITY
P/m2·s·sr·MeV/nuc
10-2
10-3
10
100
1000
ENERGY (MeV/nuc)

FIGURE 1

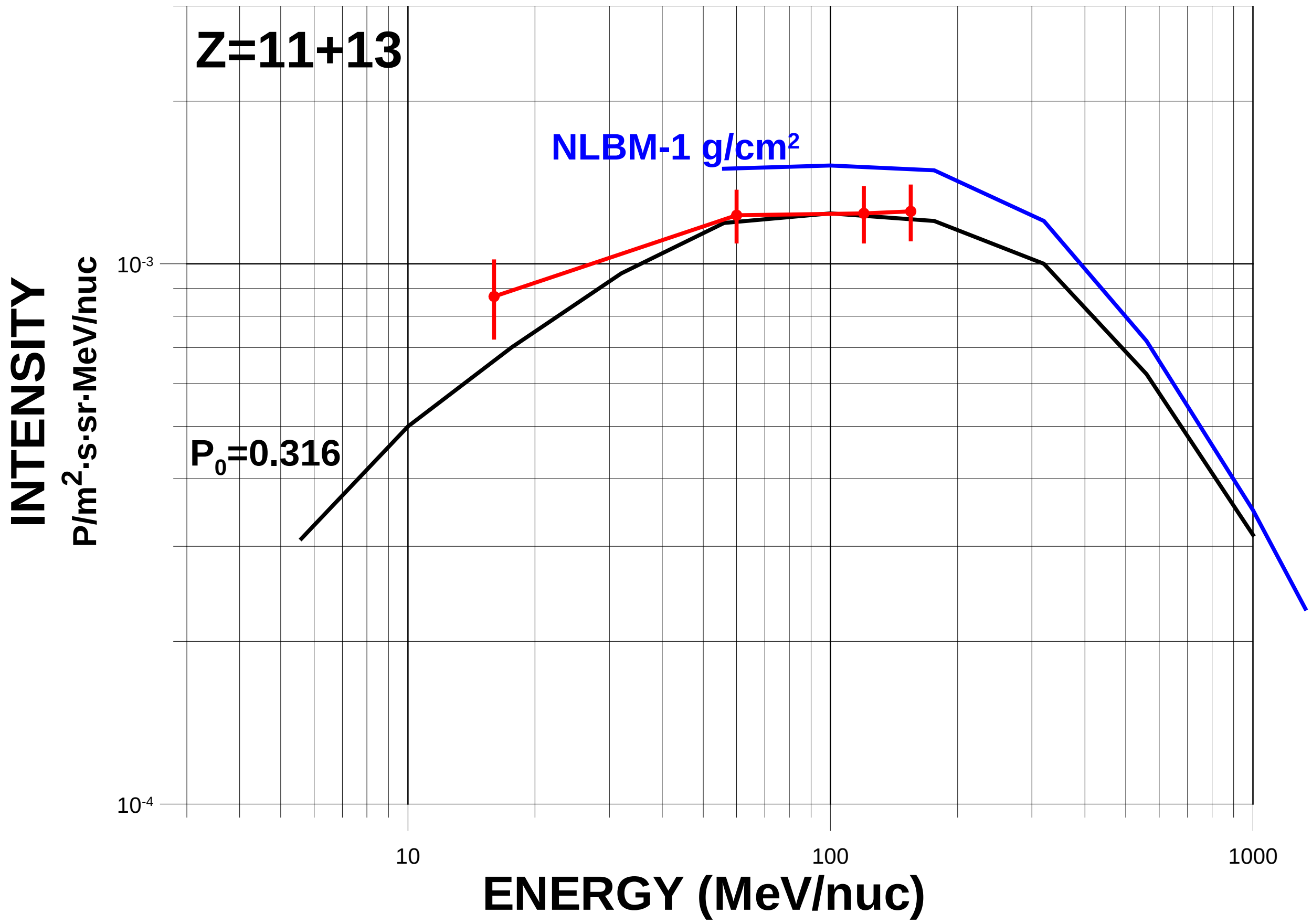
Z=11+13
NLBM-1 g/cm2
P0=0.316
INTENSITY
P/m2·s·sr·MeV/nuc
10-3
10-4
10
100
1000
ENERGY (MeV/nuc)

FIGURE 2

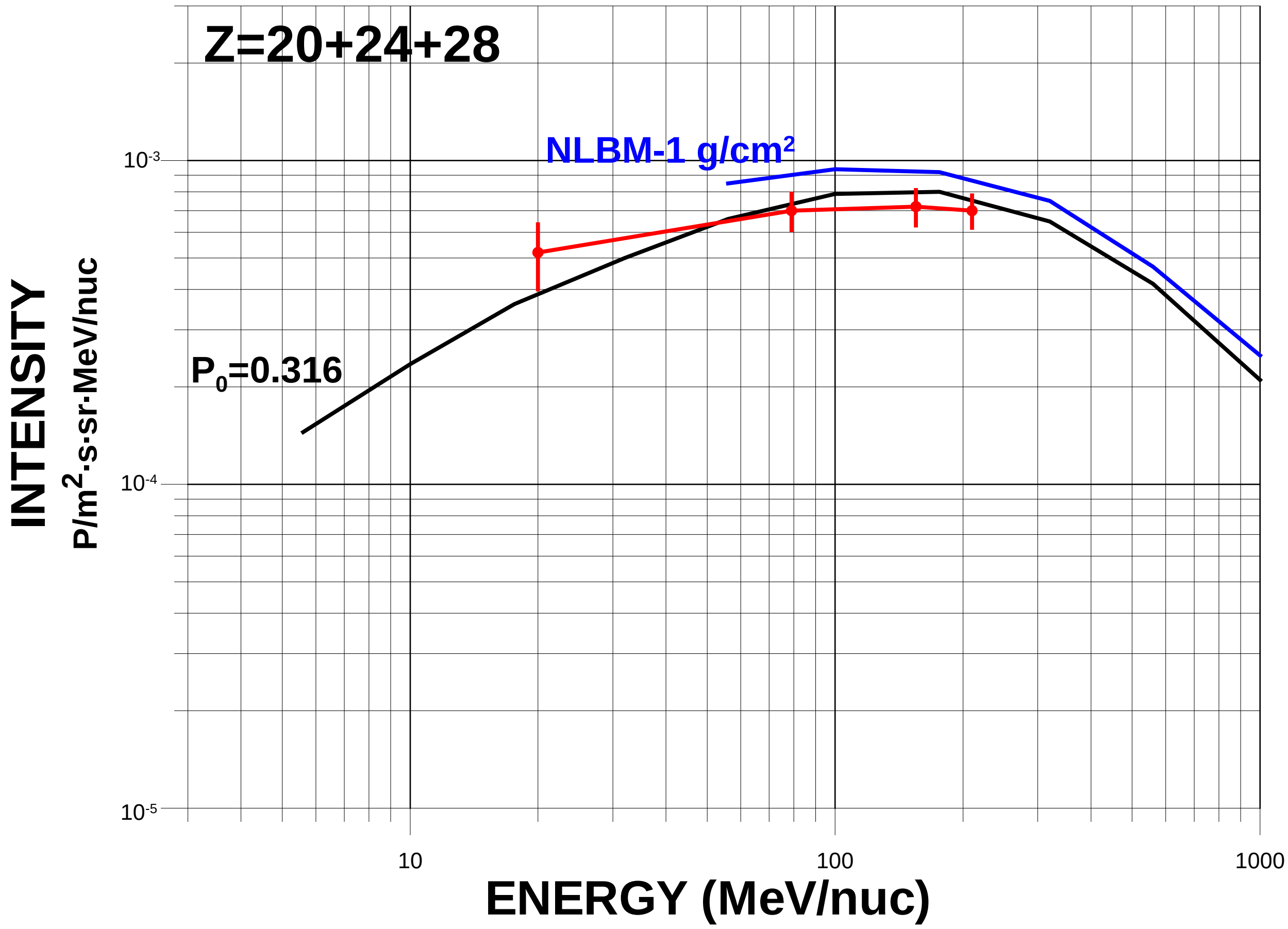
Z=20+24+28
NLBM-1 g/cm2
P0=0.316
INTENSITY
P/m2·s·sr·MeV/nuc
10-3
10-4
10-5
10
100
1000
ENERGY (MeV/nuc)

FIGURE 3

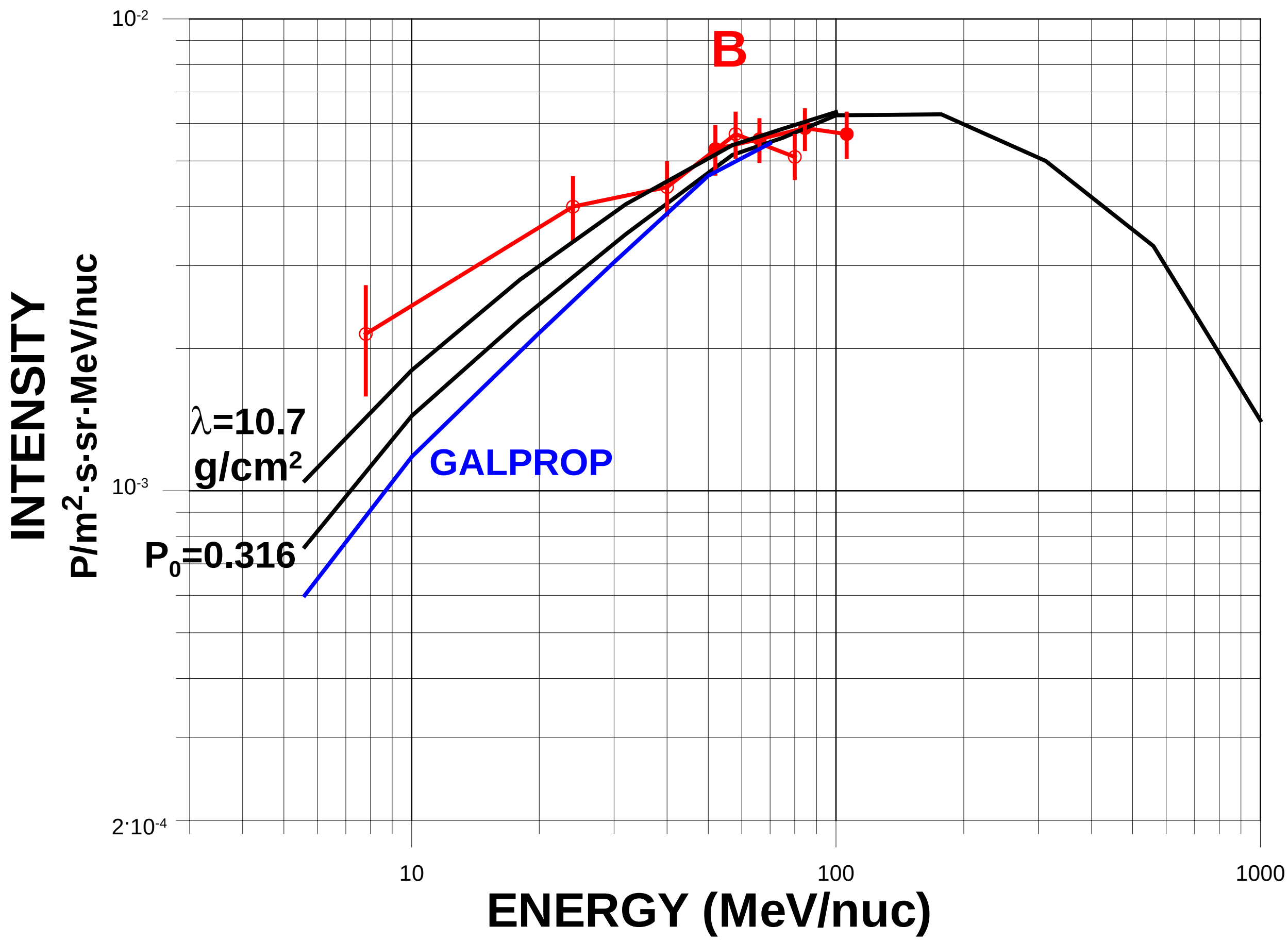

10-2
10-3
2·10-4
B
λ=10.7 g/cm2
P0=0.316
GALPROP
INTENSITY
P/m2·s·sr·MeV/nuc
10
100
1000
ENERGY (MeV/nuc)

FIGURE 4

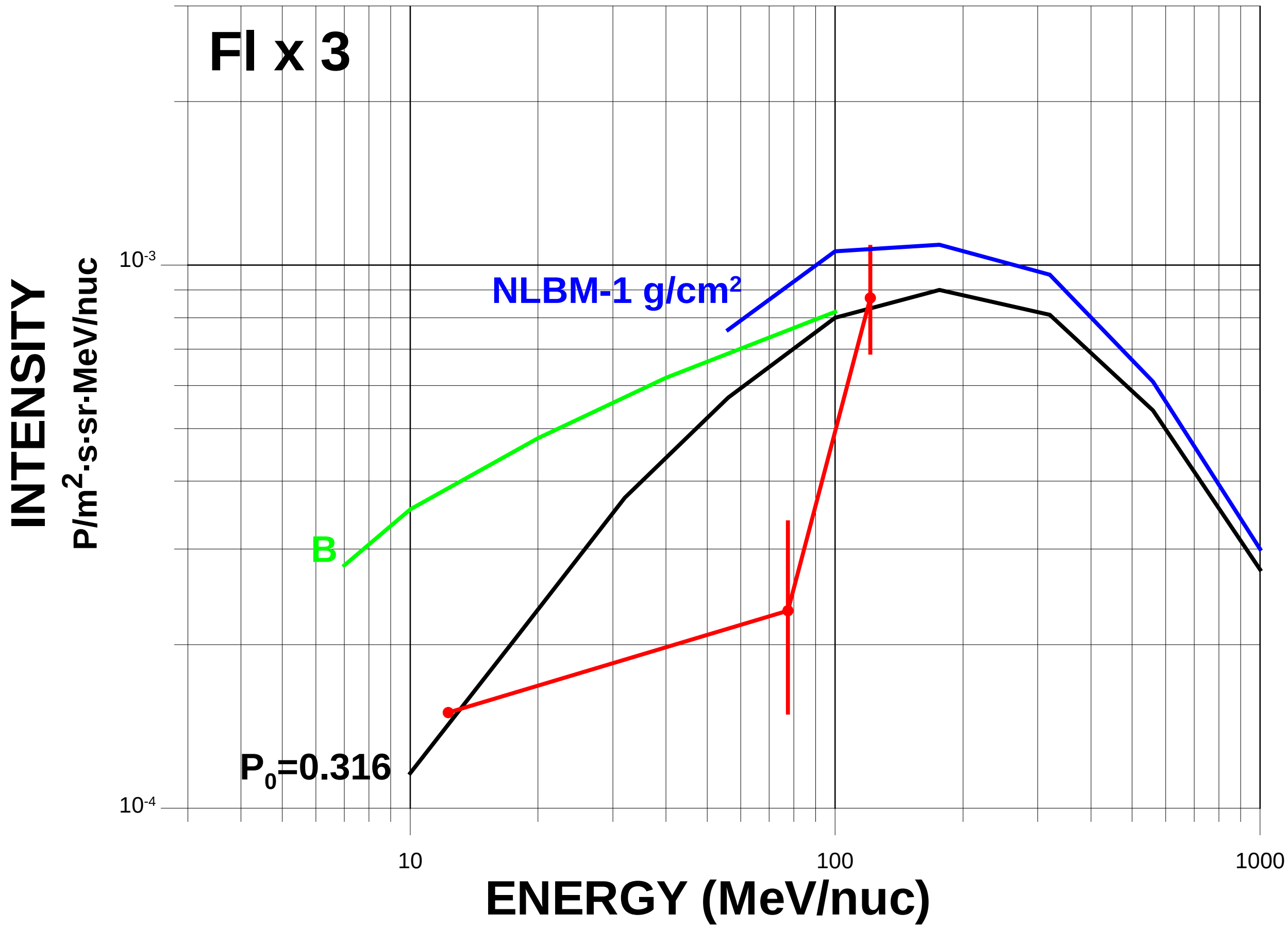

Fl x 3
INTENSITY
P/m2·s·sr·MeV/nuc
10-3
10-4
NLBM-1 g/cm2
B
P0=0.316
10
100
1000
ENERGY (MeV/nuc)

FIGURE 5

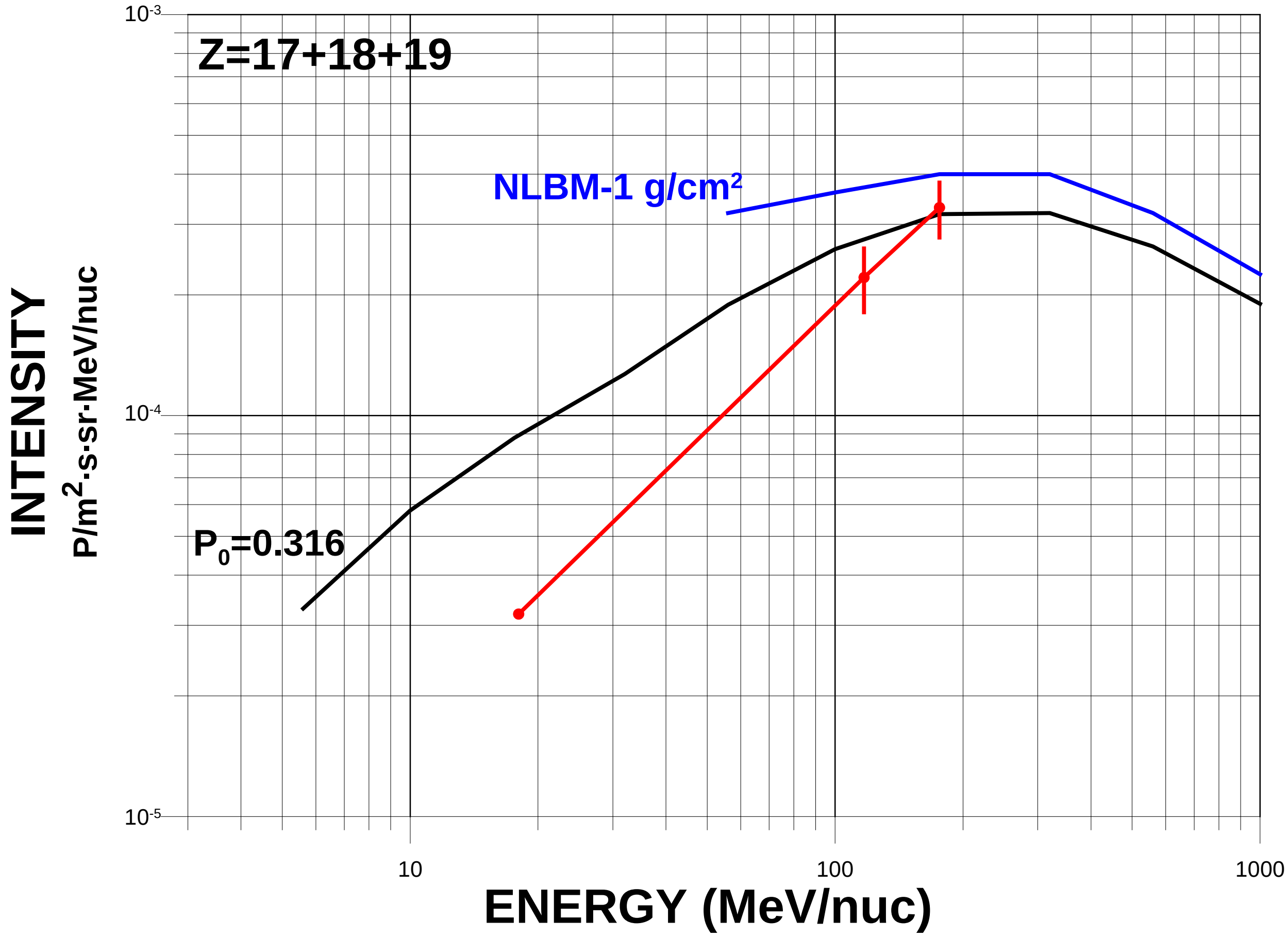

10-3
10-4
10-5
Z=17+18+19
NLBM-1 g/cm2
P0=0.316
INTENSITY
P/m2·s·sr·MeV/nuc
10
100
1000
ENERGY (MeV/nuc)

FIGURE 6

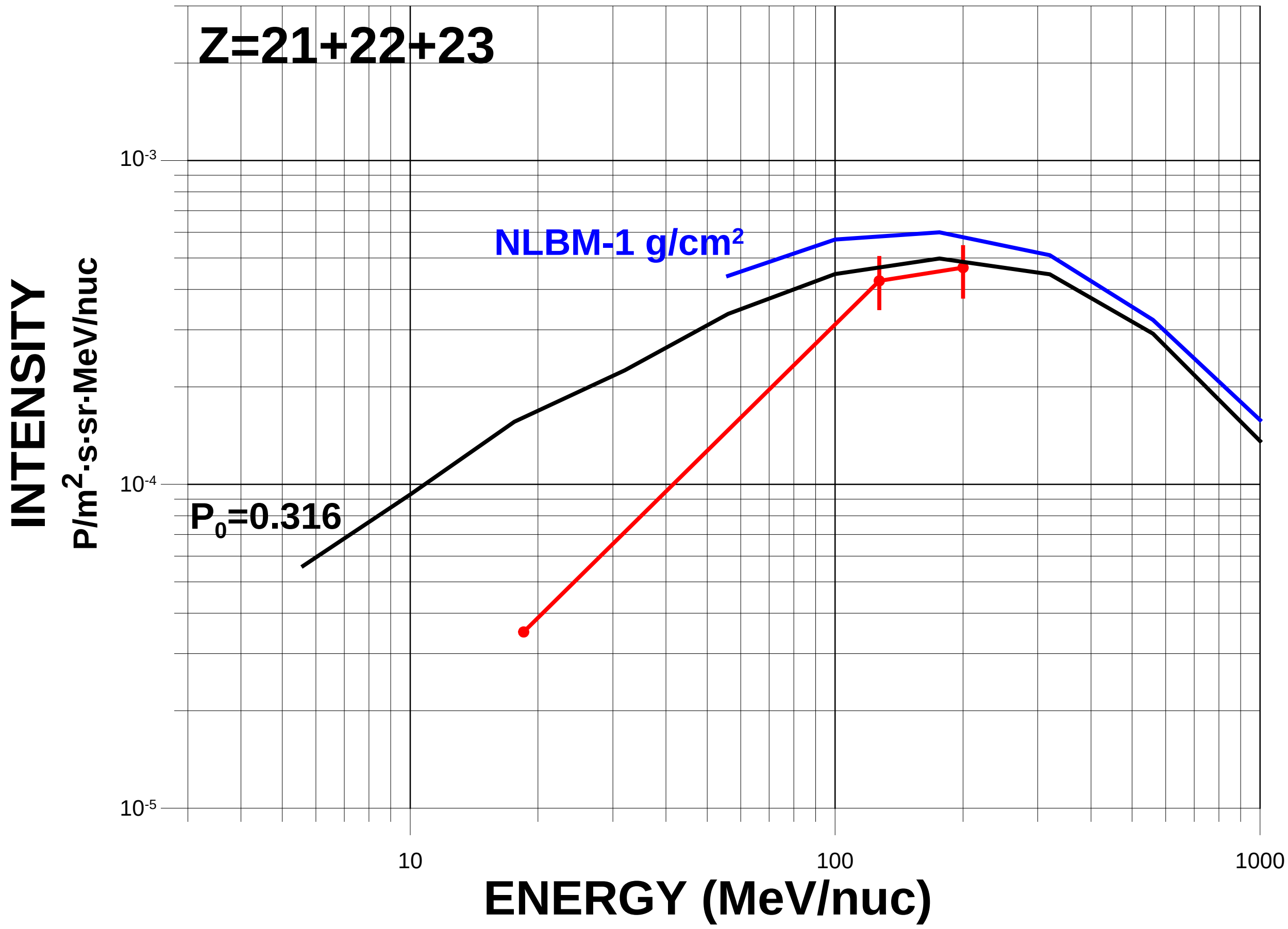
Z=21+22+23
NLBM-1 g/cm2
P0=0.316
INTENSITY
P/m2·s·sr·MeV/nuc
10-3
10-4
10-5
10
100
1000
ENERGY (MeV/nuc)

FIGURE 7